\documentclass[pra,showpacs,amsmath,amssymb,amsfonts,twocolumn,nofootinbib,superscriptaddress]{revtex4}

\usepackage{graphicx}
\usepackage{dcolumn}
\def\ket #1{\vert #1\rangle}

\begin{document}

\preprint{APS/123-QED}

\title{Quantum state engineering, purification, and number resolved photon detection with high finesse optical cavities}

\author{Anne E. B. Nielsen}
\affiliation{Lundbeck Foundation Theoretical Center for
Quantum System Research, Department of Physics and Astronomy,
Aarhus University, DK-8000 {\AA}rhus C, Denmark}
\author{Christine A. Muschik}
\author{Geza Giedke}
\author{K. G. H. Vollbrecht}
\affiliation{Max-Planck--Institut f\"ur Quantenoptik,
Hans-Kopfermann-Strasse, D-85748 Garching, Germany}

\begin{abstract}
We propose and analyze a multi-functional setup consisting of high
finesse optical cavities, beam splitters, and phase shifters. The
basic scheme projects arbitrary photonic two-mode input states
onto the subspace spanned by the product of Fock states
$|n\rangle|n\rangle$ with $n=0,1,2,\ldots$. This protocol does
not only provide the possibility to conditionally generate highly entangled
photon number states as resource for quantum information protocols but also allows one to test and hence purify this type of quantum states in a communication scenario, which is of great practical importance. The scheme is especially attractive as a generalization to many modes allows for distribution and purification of entanglement in networks. In an alternative working mode, the setup allows of quantum non demolition number resolved photodetection in the optical domain.
\end{abstract}

\pacs{42.50.Dv, 42.50.Pq, 03.67.-a}
\keywords{Suggested keywords}

\maketitle

\section{Introduction}\label{Introduction}

Light plays an essential role in quantum communication and is
indispensable in most practical applications, for example quantum
cryptography. Photons are attractive carriers of quantum
information because the interactions of light with the
surroundings are normally weak, but for the same reason it is
generally difficult to prepare, manipulate, and measure quantum
states of light in a nondestructive way. Repeated interactions
provide a method to increase the effective coupling strength
between light and matter, and the backreflection of light in a
cavity thus constitutes an interesting tool, in particular,
because experiments are currently moving into the strong coupling
regime
\cite{strongcoupling1,strongcoupling2,strongcoupling3,strongcoupling4},
where coherent dynamics takes place on a faster time scale than
dissipative dynamics.

In this paper we propose a versatile setup consisting of an array
of cavities and passive optical elements (beam splitters and phase
shifters), which allows for quantum state engineering, quantum
state purification, and non-destructive number resolving photon
detection. The setup builds on two basic ingredients: The Hong-Ou-Mandel
interference effect \cite{HongOuMandel} generalized to input
pulses containing an arbitrary number of photons and the
possibility of projection onto the subspace of even or the
subspace of odd photon-number states by use of cavity quantum
electrodynamics in the strong coupling regime.

Regarding quantum state engineering, the basic setup provides a
possibility to conditionally generate photon-number
correlated states. More specifically, the setup allows us to project an arbitrary photonic two-mode input state
onto the subspace spanned by the state vectors
$|n\rangle|n\rangle$ with $n=0,1,2,\ldots$. We denote this
subspace by $S$. The scheme is probabilistic as
it is conditioned on a specific measurement outcome. The success
probability equals the norm of the projection of the input state
onto $S$ and is thus unity if the input state already lies in $S$.
In other words, the setup may be viewed as a filter
\cite{EntanglementFilter}, which removes all undesired components
of the quantum state but leaves the desired components unchanged.
We may, for example, use two independent coherent states as
input and obtain a photon-number correlated state as output.

Photon-number correlated states, for example
Einstein-Podolsky-Rosen (EPR) entangled states \cite{EPR}, are an
important resource for quantum teleportation \cite{Teleportation1,
Teleportation2, Teleportation3, Teleportation4, Teleportation5},
entanglement swapping \cite{Swapping1, Swapping2, Swapping3},
quantum key distribution \cite{QKD1, QKD2}, and Bell tests
\cite{Bell1, Bell2}. In practice, however, the applicability of
these states is hampered by noise effects such as photon losses.
Real-world applications require therefore entanglement
purification. The proposed setup is very attractive for
detection of losses and can in particular be used to purify
photon-number entangled states on site. If a photon-number
correlated state, for example an EPR state, is used as input, the
desired state passes the setup with a certificate, while states
which suffered from photon losses are detected and can be
rejected.

Photon losses are an especially serious problem in quantum
communication over long distances. It is not only a very common
source of decoherence which is hard to avoid, but also typically
hard to overcome. The on-site purification protocol mentioned
above can easily be adopted to a communication scenario such that
it allows for the purification of a photon-number correlated state
after transmission to two distant parties.

Purification of two mode entangled states has been shown
experimentally for qubits \cite{Purification1,Purification2} and
in the continuous variable (CV) regime \cite{Purification3,
Purification4}. (CV-entanglement purification is especially
challenging \cite{GaussianImp1,GaussianImp2,GaussianImp3}. Nevertheless, several proposals have been made to accomplish this task \cite{PurificationProp1,PurificationProp2,PurificationProp3, PurificationProp4,PurificationProp5,PurificationProp6}, and very recently Takahashi {\it et al.}\ succeeded in an experimental demonstration \cite{jonas}.) A special advantage of our scheme lies in the fact that it does not only allow for detection of arbitrary photon losses, but is
also applicable to many modes such that entanglement can be
distributed and purified in a network.

With a small modification, the basic setup can be used for
number resolved photon detection. The ability to detect photons in
a number resolved fashion is highly desirable in the fields of
quantum computing and quantum communication. For example, linear optics quantum computation relies crucially on
photon number resolving detectors \cite{KLM,Kok07,Obrian07}.
Moreover, the possibility to distinguish different photon-number
states allows for conditional state preparation of nonclassical
quantum states \cite{Prep1,Prep2,Prep3}, and plays a role in Bell
experiments \cite{Zukowski93} and the security in quantum
cryptographic schemes \cite{Crypto1,Crypto2}. Other applications
include interferometry \cite{Interferometry} and the
characterization of quantum light sources
\cite{LightSources1,LightSources2}.

Existing technologies for photon counting
\cite{APD,Silberhorn04,Banzek03,Cryo1,Cryo2,Cryo3,Others1,Others2,Others3,QuantumDots1,QuantumDots2,QuantumJumps,
QND1, QND2} such as avalanche photodiodes, cryogenic devices, and
quantum dots typically have scalability problems and cannot
reliably distinguish high photon numbers, destroy the quantum
state of light in the detection process, or do not work for
optical photons. Here, we present a non-destructive number
resolving photo detection scheme in the optical regime. This
quantum-non-demolition measurement of the photon number allows for
subsequent use of the measured quantum state of light. An
advantage of the counting device put forward in this work compared
to other theoretical proposals for QND measurements of photon
numbers
\cite{QNDprop1,QNDprop2,QNDprop3,QNDprop4,QNDprop5,QNDprop6} is
the ability to detect arbitrarily high photon numbers with
arbitrary resolution. The scheme is based on testing successively all possible prime factors and powers of primes and the resources needed therefore scale moderately with (width and mean of the) photon number distribution. In particular, a very precise photon number measurement can be made even for very high photon numbers by testing only few factors if the approximate photon
number is known.

The paper is structured as follows. We start with a brief
overview of the main results in Sec.~\ref{Overview}. In
Sec.~\ref{Filter}, we explain how the conditional projection onto
$S$ can be achieved and discuss some properties of the proposed
setup in the ideal limit, where the atom-cavity coupling is infinitely strong and the input pulses are infinitely long. In Sec.~\ref{Detector}, we show that a modified version of the setup can act as a non-destructive photon number resolving detector, and in Sec.~\ref{Purification}, we investigate the
possibility to use the setup to detect, and thereby filter out, losses. In Sec.~\ref{Nonideal}, we consider the significance of finite input pulse length and finite coupling strength, and we obtain a simple analytical expression for the optimal choice of input mode function for coherent state input fields. Section~\ref{Conclusion} concludes the paper.

\section{Overview and main results}\label{Overview}

\begin{figure}
\includegraphics[width=\columnwidth]{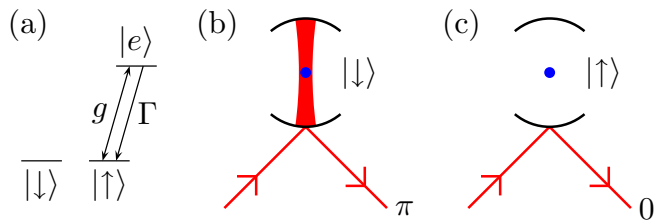}
\caption{\label{cavity}(Color online) A single atom with level
structure as shown in (a) is placed in a cavity in the strong
coupling regime. The light field, which is on resonance with the
cavity, couples the ground state level $|{\uparrow}\rangle$
resonantly to the excited state $|e\rangle$, and the state
$|e\rangle$ decays to the state $|{\uparrow}\rangle$ through
spontaneous emission at a rate $\Gamma$. (b) If the atom is in the
state $|{\downarrow}\rangle$, the incident field is not affected
by the presence of the atom, and for a sufficiently slowly varying
input pulse, the interaction with the resonant cavity changes the
phase of the light field by $\pi$. (c) If the atom is initially in
the state $|{\uparrow}\rangle$, the possibility of spontaneous
emission prevents the light field from building up inside the
cavity (provided the photon flux of the input beam is not too
high), and the incoming field is reflected from the input mirror
without acquiring a phase shift. This transformation is insensitive to the precise values of $g$, $\Gamma$, and the cavity decay rate as long as the system is in the strong coupling and weak driving regime.}
\end{figure}

The most important ingredient of the proposed setup is the possibility to use
the internal state of a single atom to control whether the phase
of a light field is changed by $\pi$ or not \cite{interaction}. The basic mechanism, which is explained in Fig.~\ref{cavity}, has
several possible applications, including preparation of
superpositions of coherent states \cite{cat}, continuous two-qubit
parity measurements in a cavity quantum electrodynamics network
\cite{parity}, and low energy switches \cite{switch}. Concerning the
experimental realization, basic ingredients of the scheme such as
trapping of a single atom in a strongly coupled cavity and preparing of
the initial atomic state have been demonstrated experimentally in \cite{reichelreadout}, where the decrease in cavity field intensity for an atom in the state $|{\uparrow}\rangle$ compared to the case of an atom in the state $|{\downarrow}\rangle$ is used to subsequently measure the state of the atom. State preparation and readout for a single atom in a cavity have also been demonstrated in \cite{rempereadout}. Another promising candidate for an experimental realization is circuit quantum electrodynamics. See, for instance, \cite{circuitQED} for a review.

The generation of quantum superposition states can be achieved as
follows. The atom is initially prepared in the state
$(|{\uparrow}\rangle+|{\downarrow}\rangle)/\sqrt{2}$, and the
input field is chosen to be a coherent state $|\alpha\rangle$.
After the interaction, the combined state of the atom and the
light field is proportional to
$|\alpha\rangle|{\uparrow}\rangle+|-\alpha\rangle|{\downarrow}\rangle\propto
(|\alpha\rangle+|-\alpha\rangle)(|{\uparrow}\rangle+|{\downarrow}\rangle)
+(|\alpha\rangle-|-\alpha\rangle)(|{\uparrow}\rangle-|{\downarrow}\rangle)$,
and a measurement of the atomic state in the basis
$\ket{\pm}=(|{\uparrow}\rangle\pm|{\downarrow}\rangle)/\sqrt{2}$
projects the state of the light field onto the even
$|\alpha\rangle+|-\alpha\rangle$ or the odd
$|\alpha\rangle-|-\alpha\rangle$ superposition state. More
generally, the input state $\sum_nc_n|n\rangle$, where $|n\rangle$
is an $n$-photon Fock state, is transformed into the output state
$\sum_n(1/2\pm(-1)^n/2)c_n|n\rangle$, i.e., the input state is
conditionally projected onto either the subspace spanned by all
even photon-number states or the subspace spanned by all odd
photon-number states without destroying the state.

With this tool at hand, we can project an arbitrary two-mode input
state onto the subspace $S=\textrm{span}(\ket{n}\ket{n})$,
$n=0,1,2,\ldots$. If two modes interfere at a 50:50 beam splitter, a state of form $|n\rangle|n\rangle$ is transformed into a superposition of products of even photon-number states. If we apply a 50:50 beam splitter operation to the input state, project both of the resulting modes conditionally on the subspace of even photon-number states, and apply a second 50:50 beam splitter
operation, the input state is thus unchanged if it already lies in
$S$, but most other states will not pass the measurement test. To
remove the final unwanted components, we apply opposite phase
shifts to the two modes (which again leaves $\ket{n}\ket{n}$
unchanged) and repeat the procedure (as shown in Fig.~\ref{setup}). For an appropriate choice of phase shifts, the
desired state is obtained after infinitely many repetitions. In
practice, however, a quite small number is typically sufficient.
If, for instance, the input state is a product of two coherent states $|\alpha\rangle|\alpha\rangle$ with $|\alpha|^2=4$, the fidelity of the projection is $0.573$ for one unit, $0.962$ for two units, and $0.999998$ for three units. The scheme is easily generalized to an $M$ mode input
state. In this case, we first project modes 1 and 2 on $S$, modes
3 and 4 on $S$, etc, and then project modes 2 and 3 on $S$, modes
4 and 5 on $S$, etc.

The setup can also be used as a device for photon number resolving
measurements if the phases applied between the light-cavity
interactions are chosen according to the new task. Each photon-number state $|n\rangle$ sent through the array leads to a
characteristic pattern of atomic states. As explained in section
\ref{DestructiveScheme}, one can determine the photon number of an
unknown state by testing the prime factors and powers of primes in
the range of interest in subsequent parts of the array. The scheme
scales thereby moderately in the resources. Three cavity pairs
suffice for example for detecting any state which is not a
multiple of three with a probability of $93.75\%$. However, in
this basic version of the counting scheme, the tested photon state
may leave each port of the last beam splitter with equal
probability. Deterministic emission of the unchanged quantum state
of light into a single spatial mode is rendered possible if
we allow atoms in different cavities to be entangled before the interaction with the field (see section \ref{QNDScheme}). More generally, the proposed scheme allows to determine the difference in photon numbers of two input beams without changing the photonic state.

The correlations in photon number between the two modes of states
in $S$ facilitate an interesting possibility to detect photon
losses. To this end the state is projected onto $S$ a second time. If photon loss has occurred, the state is most likely orthogonal to $S$, in which case we obtain a measurement outcome, which is not the one we require in order to accept the projection as successful. On the other hand, if photon loss has
not occurred, we are sure to get the desired measurement outcome.
We note that loss of a single photon can always be detected by
this method, and the state can thus be conditionally recovered
with almost perfect fidelity if the loss is sufficiently small. We
can improve the robustness even further, if we use an $M$-mode
state. It is then possible to detect all losses of up to $M-1$
photons, and even though it is $M$ times more likely to lose one
photon, the probability to lose one photon from each mode is
approximately $(Mp)^M$, where $p$ is the probability to lose one
photon from one mode and we assume $Mp\ll1$. In a situation where
many photon losses are to be expected, this procedure allows one
to obtain photon-number correlated states with high fidelity,
although with small probability.

We can also distribute the modes of a photon-number correlated
state to distant parties, while still checking for loss, provided
we send at least two modes to each party. As the proposed scheme
can be used as a filter prior to the actual protocol it has an
important advantage compared to postselective schemes. If the
tested entangled state is for example intended to be used for
teleportation, the state to be teleported is not destroyed in the
course of testing the photon-number correlated resource state.

The dynamics in Fig.~\ref{cavity} requires strong coupling, a
sufficiently slowly varying mode function of the input field, and
a sufficiently low flux of photons. To quantify these
requirements, we provide a full multi-mode description of the
interaction of the light with the cavity for the case of a
coherent state input field in the last part of the paper. We find
that the single atom cooperativity parameter should be much larger
than unity, the mode function of the input field should be long
compared to the inverse of the decay rate of the cavity, and the
flux of photons in the input beam should not significantly exceed
the rate of spontaneous emission events from an atom having an average
probability of one half to be in the excited state. We also derive
the optimal shape of the mode function of the input field
(Eq.~\eqref{mode}), when the mode function is only allowed to be
nonzero in a finite time interval.

\section{Nondestructive projection onto photon-number correlated states}\label{Filter}

\begin{figure*}
\includegraphics[width=\textwidth]{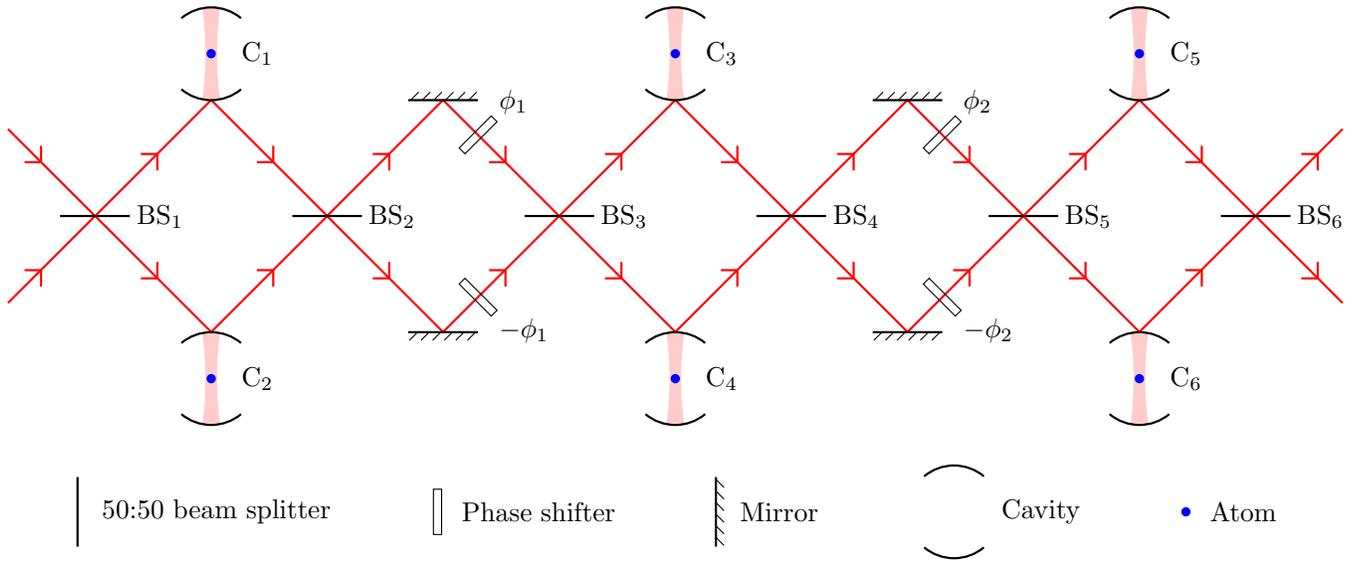}
\caption{\label{setup}(Color online) The first three units of the proposed setup to conditionally project an arbitrary two-mode input state onto the subspace spanned by the state vectors $|n\rangle|n\rangle$, $n=0,1,2,\ldots$. All atoms are prepared in the state $(|{\uparrow}\rangle+|{\downarrow}\rangle)/\sqrt{2}$ before the interaction with the field, and the desired projection occurs in the limit of infinitely many units conditioned on all atoms being in the state $(|{\uparrow}\rangle+|{\downarrow}\rangle)/\sqrt{2}$ after the interaction. As explained in the text, a small number of units will typically suffice in practice. For later reference we label the beam splitters as BS$_i$ and the cavities as C$_i$.}
\end{figure*}

The proposed setup for projection of an arbitrary two-mode input
state onto $S$ is sketched in Fig.~\ref{setup}. We denote the
field annihilation operators of the two input modes by $\hat{a}$
and $\hat{b}$, respectively. The total transformation corresponding to one of the units consisting of a beam splitter, a set of cavities, and a second beam splitter, conditioned on both atoms being measured in the state $|+\rangle$ after the interaction, is given by the operator $U^\dag PU$, where
\begin{equation}\label{UBS}
U=\exp\left[\frac{\pi}{4}\left(\hat{a}^\dag\hat{b}-\hat{a}\hat{b}^\dag\right)\right]
\end{equation}
and
\begin{equation}
P=\sum_{n=0}^\infty\sum_{m=0}^\infty|2n\rangle\langle2n|\otimes|2m\rangle\langle2m|.
\end{equation}
As explained above, the Hong-Ou-Mandel effect ensures that $U^\dag PU|n\rangle|n\rangle=|n\rangle|n\rangle$, while most other possible components of the input state are removed through the conditioning, for instance all components $|n\rangle|m\rangle$ with $n+m$ odd. There are, however, a few exceptions, since all states of the form $U^\dag|2n\rangle|2m\rangle$, $n=0,1,2,\ldots$, $m=0,1,2,\ldots$, are accepted. The phase shifts between the $U^\dag PU$ units are represented by the operator
\begin{equation}\label{Uphi}
U_\phi=\exp\left[i\phi\left(\hat{a}^\dag\hat{a}-\hat{b}^\dag\hat{b}\right)\right],
\end{equation}
which leaves states of the form $|n\rangle|n\rangle$ unchanged, while states of the form $|n\rangle|m\rangle$ with $n\neq m$ acquire a phase shift.

For a setup containing $N+1$ units, the complete conditional transformation is thus represented by the operator
\begin{eqnarray}
\hat{O}_N&=&U^\dag PUU_{\phi_N}U^\dag PU\cdots U_{\phi_2}U^\dag PU
U_{\phi_1}U^\dag PU\\
&=&U^\dag PU\prod_{i=1}^N\cos[\phi_i(\hat{a}^\dag\hat{a}-\hat{b}^\dag\hat{b})]\label{ON},
\end{eqnarray}
where $U^\dag PU$ in the last line commutes with the product of
cosines. For $N \rightarrow \infty$, the product of cosines
vanishes for all components of the input state with different
numbers of photons in the two modes if, for instance, all the
$\phi_i$'s are chosen as an irrational number times $\pi$. We note
that even though we here apply the two-mode operators one after
the other to the input state corresponding to successive
interactions of the light with the different components of the
setup, the result is exactly the same if the input pulses are
longer than the distance between the components such that
different parts of the pulses interact with different components
at the same time. The only important point is that the state of
the atoms is not measured before the interaction with the light
field is completed. The setup using an array of cavities as in Fig.~\ref{setup} may thus be very compact even though the pulses are required
to be long. (Note that it would also be possible to use a single pair of
cavities and atoms repeatedly in a fold-on type of experiment; however
in that case the compactness would be lost due to the need for long
delay lines necessary to measure and re-prepare the atoms before they
are reused.)

A natural question is how one should optimally choose the angles $\phi_i$ to approximately achieve the projection with a small number of units. To this end we define the fidelity of the projection
\begin{equation}\label{FN}
F_N=\frac{|\langle\psi_N|\psi_\infty\rangle|^2}
{\langle\psi_N|\psi_N\rangle\langle\psi_\infty|\psi_\infty\rangle}
=\frac{\langle\psi_\infty|\psi_\infty\rangle}
{\langle\psi_N|\psi_N\rangle}
\end{equation}
as the overlap between the unnormalized output state $|\psi_N\rangle=\hat{O}_N|\psi_{\textrm{in}}\rangle$ after $N+1$ units and the projection $|\psi_\infty\rangle$ of the input state $|\psi_{\textrm{in}}\rangle$ onto the subspace $S$. The last equality follows from the fact that $|\psi_N\rangle=|\psi_\infty\rangle+|\psi_\bot\rangle$, where $|\psi_\bot\rangle$ lies in the orthogonal complement of $S$. Maximizing $F_N$ for a given $|\psi_{\textrm{in}}\rangle=\sum_n\sum_mc_{nm}|n\rangle|m\rangle$ thus corresponds to minimizing
\begin{multline}
\langle\psi_N|\psi_N\rangle=\sum_{n=0}^\infty\sum_{m=0}^\infty c_{nm}
\prod_{i=1}^N\cos^2[\phi_i(n-m)]\\
\times\langle\psi_{\textrm{in}}|U^\dag PU|n\rangle|m\rangle,
\end{multline}
i.e., we would like to find the optimal solution of
\begin{multline}
\frac{\partial\langle\psi_N|\psi_N\rangle}{\partial\phi_j}=
-\sum_{n=0}^\infty\sum_{m=0}^\infty c_{nm}\sin[2\phi_j(n-m)](n-m)\\
\times\prod_{\substack{i=1\\i\neq j}}^N\cos^2[\phi_i(n-m)]
\langle\psi_{\textrm{in}}|U^\dag PU|n\rangle|m\rangle=0.
\end{multline}
A set of solutions valid for any input state can be obtained by requiring $\sin[2\phi_j(n-m)]\prod_{i\neq j}\cos^2[\phi_i(n-m)]=0$ for all even values of $n-m$ (note that $U^\dag PU|n\rangle|m\rangle=0$ for $n+m$ odd). Within this set the optimal solution is $\phi_j=2^{-j}\times\pi/2$. It is interesting to note that choosing one of the angles to be $2^{-j}\times\pi/2$, $j\in\{1,2,\ldots,N\}$, all terms with $n-m=2^j\times(\pm1,\pm3,\pm5,\ldots)$ are removed from the input state according to \eqref{ON}. When all angles are chosen according to $\phi_j=2^{-j}\times\pi/2$, it follows that $|\psi_N\rangle$ only contains terms with $n-m=q2^{(N+1)}$, $q=0,\pm1,\pm2,\ldots$, which may be a useful property in practical applications of the scheme. Even though this is not necessarily optimal with respect to maximizing $F$ for a particular choice of input state, we thus use the angles $\phi_j=2^{-j}\times\pi/2$ in the following, except for one important point: If the input state satisfies the symmetry relations $c_{nm}=c_{mn}$, it turns out that the operator $U^\dag PU$ by itself removes all terms with $n-m=\pm2,\pm6,\pm10,\ldots$, i.e., we can choose the angles as $\phi_j=2^{-j}\times\pi/4$, and $|\psi_N\rangle$ only contains terms with $n-m=q2^{(N+2)}$, $q=0,\pm1,\pm2,\ldots$. For $N=2$, for instance, only terms with $n-m=0,\pm16,\pm32,\ldots$ contribute.

\begin{figure}
\includegraphics[width=\columnwidth]{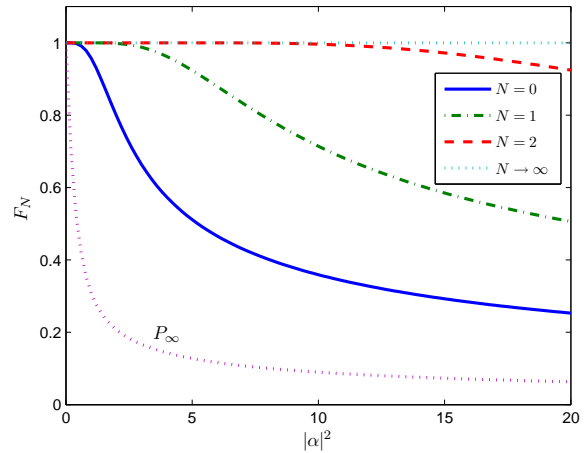}
\caption{\label{fidelity}(Color online) Fidelity (Eq.~\eqref{FN}) as a function of the expectation value of the number of photons in one of the input modes for $|\psi_{\textrm{in}}\rangle=|\alpha\rangle|\alpha\rangle$ and setups with one, two, three, and infinitely many units. The angles are chosen as $\phi_j=2^{-j}\times\pi/4$. The dotted line labeled $P_{\infty}$ is the probability in the limit of infinitely many units to actually obtain the required measurement outcome, i.e., all atoms in $(|{\uparrow}\rangle+|{\downarrow}\rangle)/\sqrt{2}$ after the interaction with the field.}
\end{figure}

In Fig.~\ref{fidelity}, we have chosen the input state to be a product of two coherent states with amplitude $\alpha$ and plotted the fidelity \eqref{FN} as a function of $|\alpha|^2$ for different numbers of units of the setup. Even for $|\alpha|^2$ as large as 10, the fidelity is still as high as 0.9961 for $N=2$, and the required number of units is thus quite small in practice. The figure also shows the success probability
\begin{equation}
P_N=\langle\psi_N|\psi_N\rangle
\end{equation}
for $N\rightarrow\infty$. For $|\alpha|^2=10$, for instance, one should repeat the experiment about 11 times on average before the desired measurement outcome is observed.

\section{Photon number resolving measurement}\label{Detector}

In this section, we show how a photon number measurement can be
implemented using a modified version of the setup introduced in
the previous section. The key idea is explained in subsection
\ref{DestructiveScheme} where we describe the basic photo-counting
scheme. In subsection \ref{QNDScheme}, this protocol is extended
to allow for a QND measurement of photon numbers.

\subsection{Number resolving detection scheme}\label{DestructiveScheme}

In the following, we analyze the setup shown in Fig.~\ref{setup}
when the input is a product of an $n$-photon Fock state in the lower input beam and a vacuum state in the upper input beam. Since the setup
contains a series of beam splitters, it will be useful to define
$\hat{A}=(\hat{a}^\dag-\hat{b}^\dag)/\sqrt{2}$ and
$\hat{B}=(\hat{a}^\dag+\hat{b}^\dag)/\sqrt{2}$, such that
$\hat{a}^\dag\ket{0}\rightarrow\hat{A}\ket{0}$ and
$\hat{b}^\dag\ket{0}\rightarrow\hat{B}\ket{0}$ at beam splitters
BS$_1$, BS$_3$, BS$_5$, $\ldots$, and
$\hat{A}\ket{0}\rightarrow\hat{a}^\dag\ket{0}$ and
$\hat{B}\ket{0}\rightarrow\hat{b}^\dag\ket{0}$ at beam splitters
BS$_2$, BS$_4$, BS$_6$, $\ldots$.

As before, all atoms are initially prepared in the state
$\ket{+}$ and will after the interaction with the field be measured in the
$\ket{\pm}$ basis. When we start with an $n$ photon state, there are only two possible outcomes of the measurement of the atoms in the cavities labeled C$_1$ and C$_2$ in Fig.~\ref{setup} dependent on whether $n$ is even or odd. In the even case, the two atoms can only be in $\ket{++}$ or $\ket{--}$ and in the odd case $\ket{-+}$ or $\ket{+-}$. To handle the odd and even case at the same
time we denote $\ket{++},\ket{-+}$ as $\ket{B_+}$ and
$\ket{--},\ket{+-}$ as $\ket{B_-}$. A measurement of $\ket{B_+}$
indicates an even number of photons in the $\hat{b}$-beam, resp.,
$\ket{B_-}$ an odd number of photons in the $\hat{b}$-beam.

We start with the state $\ket{n}\ket{0}=\frac{1}{\sqrt{n!}}(\hat{a}^\dag)^n\ket{0}\ket{0}$ as input. After the beam splitter BS$_1$, the state has changed into $\frac{1}{\sqrt{n!}}\left(\frac{\hat{a}^\dag-\hat{b}^\dag}{\sqrt{2}}\right)^n\ket{0}\ket{0}$
and interacts with the two atoms in the cavities C$_1$ and C$_2$. By measuring the atoms in the $\ket{\pm}$ basis, the state is projected into the subspace of an even or odd number of photons in the $\hat{b}$ path. The photon state after the measurement can be written as
$\ket{b_\pm}:=\frac{1}{\sqrt{2\
n!}}\left[\left(\frac{\hat{a}^\dag-\hat{b}^\dag}
{\sqrt{2}}\right)^n \pm\left(\frac{\hat{a}^\dag+\hat{b}^\dag}
{\sqrt{2}}\right)^n\right]\ket{0}\ket{0}=\frac{1}{\sqrt{2\
n!}}(\hat{A}^n \pm\hat{B}^n)\ket{0}\ket{0}$, where
$\ket{b_+}$($\ket{b_-}$) is the state with an even (odd) number of
$\hat{b}$ photons and corresponds to the measurement result
$\ket{B_+}$ ($\ket{B_-}$).  Note that this first measurement result is
completely random.

After BS$_2$ the state simplifies to
$\frac{1}{\sqrt{2\ n!}}[(\hat{a}^\dag)^n
\pm(\hat{b}^\dag)^n]\ket{0}\ket{0}$. Now due to the phase shifters the two modes pick up a relative phase of $2 \phi_1 n$ so that the state is given by $\frac{1}{\sqrt{2\ n!}}[e^{i \phi_1 n}(\hat{a}^\dag)^n \pm e^{-i \phi_1 n}(\hat{b}^\dag)^n]\ket{0}\ket{0}$. Finally, after BS$_3$ we have the state $\frac{1}{\sqrt{2\ n!}} (e^{i\phi_1 n}\hat{A}^n \pm e^{-i\phi_1 n}\hat{B}^n)\ket{0}\ket{0}$, which is equal to
$\frac{1}{2\sqrt{2\ n!}}(e^{i\phi_1 n} \pm e^{-i\phi_1 n})(\hat{A}^n +
\hat{B}^n)\ket{0}\ket{0}+\frac{1}{2\sqrt{2\ n!}}(e^{i\phi_1
n} \mp e^{-i \phi_1 n})(\hat{A}^n - \hat{B}^n)\ket{0}\ket{0}$. This can also be rewritten as $(e^{i\phi_1 n}\pm e^{-i\phi_1 n})/2\ket{b_+}+(e^{i \phi_1
n} \mp e^{-i \phi_1 n})/2\ket{b_-}$. So the result of measuring the state of the atoms in cavities C$_3$ and C$_4$ will be $\ket{B_\pm}$
with probability $p_+=\cos(\phi_1n)^2$
and $\ket{B_\mp}$ with probability $p_-=\sin(\phi_1n)^2$. Since the state is again projected into one of the two states $\ket{b_\pm}$ we can repeat exactly the same calculations for all following steps.

Whereas the first measurement result was completely random, all
following measurement results depend on $n$ and the previous
measurement outcome, i.e.,  with probability $p_i=\cos(\phi_i
n)^2$ the $(i+1)$-th measurement result is the same as the $i$-th
result, and with probability $\sin(\phi_i n)^2$ the measurement result
changes and the state changes from $\ket{b_\pm}$ to $\ket{b_\mp}$.
If the number of units is infinite (or sufficiently large) and we have
chosen all phases equal as $\phi$, then we can guess from the
relative frequency with which the measurement result has switched
between $\ket{B_+}$ and $\ket{B_-}$ the number of photons with
arbitrary precision for all photon numbers $n<\frac{\pi}{\phi}$. $n\approx \arccos(\sqrt{f})/\phi$, where $f=N_\textrm{same}/(N_\textrm{same}+N_\textrm{different})$ and $N_\textrm{same}$ ($N_\textrm{different}$) is the number of cases, where the measurement outcome is the same (not the same) as the previous measurement outcome.

Measuring this relative frequency with a fixed small phase is not
the optimal way to get the photon number. We propose instead the
following. Let us use a setup with a total of $N+1$ units and
choose the phases to be $\phi_i=2^{i-1} \pi/n_0$,
$i=1,2,\ldots,N$, for an arbitrarily chosen value of $n_0 \in
\mathbb{N}$ and let us calculate the probability $p(n)$ that the
measurement results are all the same,
\begin{equation}\label{pn}
p(n)=\prod_{i=1}^N p_i=\prod_{i=0}^{N-1} \cos\left(\frac{2^i\pi}{n_0}n\right)^2.
\end{equation}
This probability is equal to one for all photon numbers that are a multiple of $n_0$ and goes to zero otherwise in the limit of infinitely many units of the setup. This way we can measure whether the photon number is a multiple of $n_0$.

\begin{figure}
\includegraphics[width=\columnwidth]{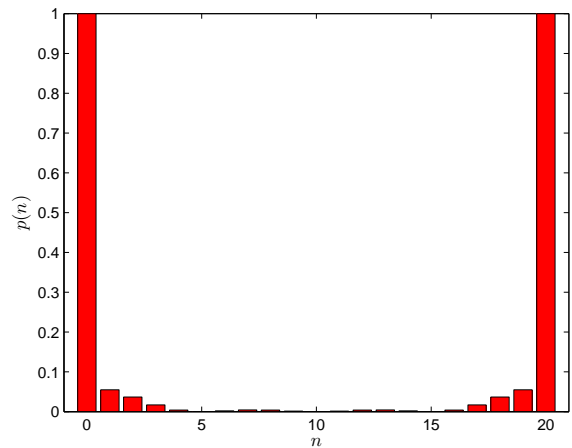}
\caption{\label{20photo}(Color online) Probabilities $p(n)$ (Eq.~\eqref{pn}) for $n_0=20$ and $N+1=5$ resulting in $q=94.49\%$ (Eq.~\eqref{q}). For
$N+1=7$, $q=99.66\%$.}
\end{figure}

\begin{figure}
\includegraphics[width=\columnwidth]{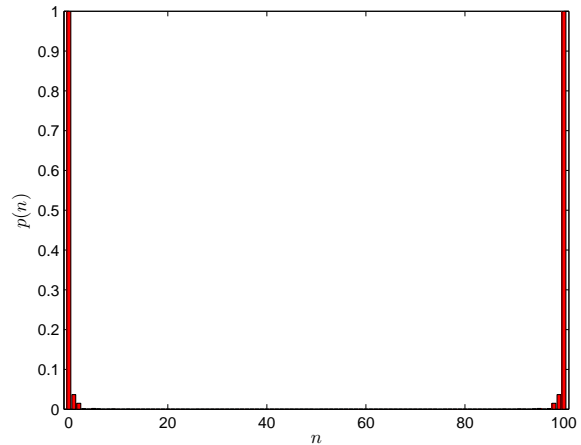}
\caption{\label{100photo}(Color online) Probabilities $p(n)$ for $n_0=100$ and
$N+1=8$ resulting in $q=96.33\%$. For $N+1=10$, $q=99.95\%$.}
\end{figure}

For example, for $n_0=3$ and $N+1=3$ we detect any state which is not a
multiple of three with a probability of at least $q=93.75\%$, resp., $q=99.61\%$ for $N+1=5$, where
\begin{equation}\label{q}
q:=1-\max_{n\neq0,n_0,2n_0,\ldots}p(n).
\end{equation}
For $n_0=4$, already $N+1=3$ is sufficient to achieve $q=100\%$. For $n_0=5$ and $N+1=3$ we have $q=93.75\%$, which increases to $q=99.61\%$ for $N+1=5$. In Fig.~\ref{20photo} and \ref{100photo} we have shown $p(n)$ for $n_0=20$ and
$n_0=100$. The number $N$ needed to get a good result typically
scales logarithmical with $n_0$, e.g. for $n_0=1000$ already
$N+1=11$ is enough to reach $q=99.95\%$.

Given an unknown state we can test all possible prime factors and powers of primes to identify the exact photon number. If, e.g., we have a state consisting of $0$ to $10$ photons the following factors have to be tested $n_0=2,3,4,5,7,8,9$ (where 2 does not need to be checked separately). 24 measurement results are sufficient to test all factors with a probability over $99\%$.

If we require reliable photon number counting for $n$ ranging from
$0$ to $n_\textrm{max}$, for large $n_\textrm{max}$, all primes
and power of primes that are smaller than $n_\textrm{max}$ need to
be tested. This number can be bounded from above by
$n_\textrm{max}$. All $n_\textrm{max}$ tests are required to work
with high probability. To this end each single test needs to
succeed with a probability better than $q\geq
1-1/n_\textrm{max}$. It can be checked numerically that this is
the case if  $N=2 \log(n_\textrm{max})$, leading to a photon
counting device with reliable photon detection up to
$n_\textrm{max}$ using an array consisting of less than $2
n_\textrm{max} \log(n_\textrm{max})$ basic units.

Note that this setup does not destroy the photonic input state but
changes $\ket{n}\ket{0}$ randomly into $\frac{1}{\sqrt{2}}[\ket{n}\ket{0}\pm\ket{0}\ket{n}]$, i.e., the photons
leave the setup in a superposition of all photons taking either
the $\hat{a}$-beam or the $\hat{b}$-beam. This can not be changed
back into $\ket{n}\ket{0}$ by means of passive optical elements.
The output state - a so-called $N00N$ state - is, however, a very
valuable resource for applications in quantum information
protocols and quantum metrology \cite{NOON}.

\subsection{Non-destructive number resolving detection scheme}\label{QNDScheme}

\begin{figure}
\includegraphics[width=\columnwidth]{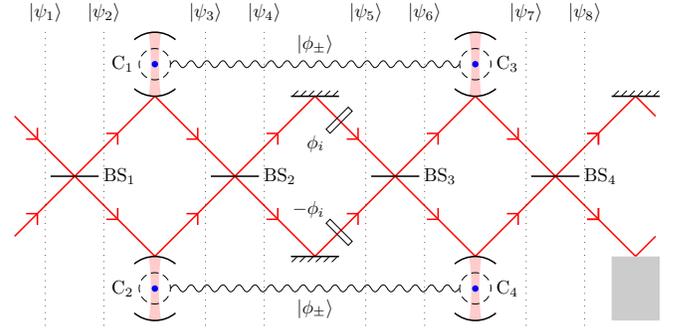}
\caption{(Color online) Basic building block of the non-demolition photon number resolving detection scheme. The dashed circles and the wavy lines indicate the entanglement between the atoms in cavities C$_1$ and C$_3$ and between the atoms in cavities C$_2$ and C$_4$. The gray box is either a cavity or a mirror. (In the latter case, we may as well send the light directly from the cavities C$_3$ and C$_4$ to the first two cavities of the next block.)} \label{block}
\end{figure}

For a non-demolition version of the photon number measurement we
use the basic building block depicted in Fig.~\ref{block}. The atoms in the two upper cavities are initially prepared in an entangled state $\ket{\phi_+}=(\ket{{\uparrow}{\uparrow}}+\ket{{\downarrow}{\downarrow}})/\sqrt{2}$, and the atoms in the lower cavities are also prepared in the state $\ket{\phi_+}$. This can, for instance, be achieved via the parity measurement scheme suggested in \cite{parity}. During the interaction the upper and the lower atoms will stay in the subspace spanned by $\ket{\phi_\pm}=(\ket{{\uparrow}{\uparrow}} \pm \ket{{\downarrow}{\downarrow}})/\sqrt{2}$. The state changes between $\ket{\phi_+}$ and $\ket{\phi_-}$ each time one of the two entangled cavities interacts with an odd number of photons. As in the
previous subsection we define $\ket{B_\pm}$ to handle the even
and the odd case at the same time. For $n$ even,
$\ket{B_+}=\ket{\phi_+}\ket{\phi_+}$ and
$\ket{B_-}=\ket{\phi_-}\ket{\phi_-}$, while in the odd case we
define $\ket{B_+}=\ket{\phi_-}\ket{\phi_+}$ and
$\ket{B_-}=\ket{\phi_+}\ket{\phi_-}$.

Using the same notation as before, the state is transformed as follows, when we go through the setup from left to right. The initial state is
\begin{equation*}
\ket{\psi_1}=\frac{1}{\sqrt{n!}} (\hat{a}^\dag)^n\ket{0}\ket{0}\ket{\phi_+}\ket{\phi_+},
\end{equation*}
and after the first beam splitter we have the state
\begin{equation*}
\ket{\psi_2}=\frac{1}{\sqrt{n!}}\hat{A}^n\ket{0}\ket{0}\ket{\phi_+}\ket{\phi_+}
=\frac{1}{\sqrt{2}}(\ket{b_+}+\ket{b_-})\ket{\phi_+}\ket{\phi_+}.
\end{equation*}
The interaction with the first two cavities leads to the state
\begin{multline*}
\ket{\psi_3}=\frac{1}{\sqrt{2}}\ket{b_+}\ket{B_+}+\frac{1}{\sqrt{2}}\ket{b_-}\ket{B_-}\\
=\frac{1}{2\sqrt{n!}}(\hat{A}^n+\hat{B}^n)\ket{0}\ket{0}\ket{B_+}\\
+\frac{1}{2\sqrt{n!}}(\hat{A}^n-\hat{B}^n)\ket{0}\ket{0}\ket{B_-}.
\end{multline*}
The second beam splitter transforms the state into
\begin{multline*}
\ket{\psi_4}=\frac{1}{2\sqrt{n!}}\big\{[(\hat{a}^\dag)^n+(\hat{b}^\dag)^n]\ket{0}\ket{0}\ket{B_+}\\
+[(\hat{a}^\dag)^n-(\hat{b}^\dag)^n]\ket{0}\ket{0}\ket{B_-}\big\}.
\end{multline*}
The two modes pick up a relative phase shift of $2 \phi_i n$ at the phase shifters so that
\begin{multline*}
\ket{\psi_5}=\frac{1}{2\sqrt{n!}}
\big\{[e^{i\phi_i n}(\hat{a}^\dag)^n
+e^{-i\phi_in}(\hat{b}^\dag)^n]\ket{0}\ket{0}\ket{B_+}\\
+[e^{i\phi_i n}(\hat{a}^\dag)^n
-e^{-i\phi_i n}(\hat{b}^\dag)^n]\ket{0}\ket{0}\ket{B_-}\big\}.
\end{multline*}
After the third beam splitter we get
\begin{multline*}
\ket{\psi_6}=\frac{1}{2\sqrt{n!}}\big[(e^{i\phi_i n}\hat{A}^n+e^{-i\phi_i n}\hat{B}^n)\ket{0}\ket{0}\ket{B_+}\\
+(e^{i\phi_i n}\hat{A}^n-e^{-i\phi_i n}\hat{B}^n)\ket{0}\ket{0}\ket{B_-}\big]\\
=\frac{1}{\sqrt{2}}\cos(\phi_i n)\ket{b_+}\ket{B_+}
+\frac{i}{\sqrt{2}}\sin(\phi_i n)\ket{b_-}\ket{B_+}\\
+\frac{i}{\sqrt{2}}\sin(\phi_i n)\ket{b_+}\ket{B_-}
+\frac{1}{\sqrt{2}}\cos(\phi_i n)\ket{b_-}\ket{B_-}.
\end{multline*}
Note now that independent of whether $n$ is even or odd, the interaction with the last two cavities turns the states $\ket{b_\pm}\ket{B_\pm}$ into $\ket{b_\pm}\ket{\phi_+}\ket{\phi_+}$ and the states $\ket{b_\pm}\ket{B_\mp}$ into $\ket{b_\pm}\ket{\phi_-}\ket{\phi_-}$. The state is thus changed to
\begin{multline*}
\ket{\psi_7}=
\frac{1}{\sqrt{2}}\cos(\phi_i n)\left(\ket{b_+}+\ket{b_-}\right)\ket{\phi_+}\ket{\phi_+}\\
+\frac{i}{\sqrt{2}}\sin(\phi_i n)\left(\ket{b_+}+\ket{b_-}\right)\ket{\phi_-}\ket{\phi_-}\\
=\frac{1}{\sqrt{n!}}\hat{A}^n\ket{0}\ket{0}
\left[\cos(\phi_i n)\ket{\phi_+}\ket{\phi_+}+i\sin(\phi_i n)\ket{\phi_-}\ket{\phi_-}\right],
\end{multline*}
and after the last beam splitter we have
\begin{multline*}
\ket{\psi_8}=\frac{1}{\sqrt{n}}(\hat{a}^\dag)^n\ket{0}\ket{0}\\
\otimes\left[\cos(\phi_i n)\ket{\phi_+}\ket{\phi_+}
+i\sin(\phi_i n)\ket{\phi_-}\ket{\phi_-}\right].
\end{multline*}
Note that the photonic modes are now decoupled from the state of the atoms. The photons will continue after the final beam splitter unchanged in $\ket{n}\ket{0}$ while the atoms still contain some information about the photon number.

Note that $\ket{\phi_+}=(\ket{++}+\ket{--})/\sqrt{2}$ and
$\ket{\phi_-}=(\ket{+-}+\ket{-+})/\sqrt{2}$. By measuring all
atoms in the $\ket{\pm}$ basis we can easily distinguish between
$\ket{\phi_\pm}$ by the parity of the measurements. The probability to obtain $\ket{\phi_+}\ket{\phi_+}$ is $\cos(\phi n)^2$, and the probability to get $\ket{\phi_-}\ket{\phi_-}$ is $\sin(\phi n)^2$. After the measurement, all photons are found in the $\hat{a}$ beam for both outcomes. The probability for measuring $\ket{\phi_+}\ket{\phi_+}$ is the same as the changing probability in the previous setup such that we can do the same with a chain of the demolition free block. If we prefer also to get the parity information in each step, we can add an additional cavity at the end of each block as shown in Fig.~\ref{block}.

More generally, the demolition free element leaves photonic input
states $\ket{\psi}=\frac{1}{\sqrt{n!q!} }
(\hat{a}^\dag)^n(\hat{b}^\dag)^q \ket{0}\ket{0}$, where $n$ photons
enter through the lower and $q$ photons enter through the upper
port, unchanged. A calculation analogous to the previous one shows
that one obtains the atomic states $\ket{\phi_+}\ket{\phi_+}$ and
$\ket{\phi_-}\ket{\phi_-}$ with probabilities $ \cos(\phi_i
(n-q))^2$ and $\sin(\phi_i (n-q))^2$ respectively. This way, we
can test for photon number differences $n-q$ in two input states
in the same fashion as for photon numbers in a single input beam
described above. Similarly, one can project two coherent input states
$\ket{\alpha}\ket{\alpha}$ onto generalized photon-number
correlated states $\sum_n c_n\ket{n}\ket{n-d}$ with fixed photon
number difference $d=0,1,2$....

In a realistic scenario we may be faced with photon losses. Both
setups have a built-in possibility to detect loss of one photon. In the first case we get the parity of the total number of photons in every single measurement of a pair of atoms. If this parity changes at some place in the chain then we know that we lost at least one photon. In the demolition free setup the valid measurement results are restricted to $\ket{\phi_+}\ket{\phi_+}$ and $\ket{\phi_-}\ket{\phi_-}$. If we measure
$\ket{\phi_-}\ket{\phi_+}$ or $\ket{\phi_+}\ket{\phi_-}$ we know that we lost a photon in between the two pairs of cavities. In addition, the optional cavity at the end of each block provides an extra check for photon loss.

\section{Filtering out losses}\label{Purification}

\begin{figure}
\includegraphics[width=\columnwidth]{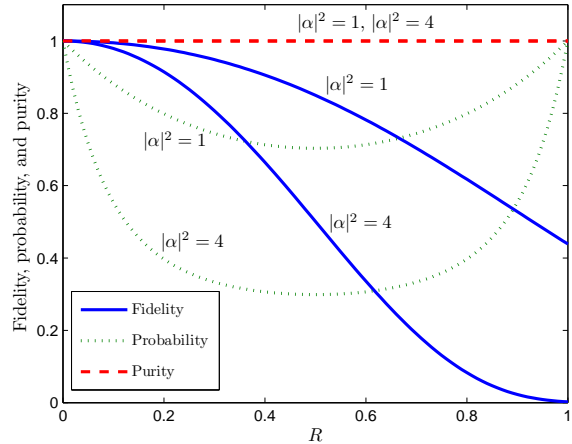}
\caption{\label{recover}(Color online) Projecting the input state $|\psi_{\textrm{in}}\rangle=|\alpha\rangle|\alpha\rangle$ onto the subspace $S$ followed by a fractional loss $R$ in both modes and a second projection onto $S$, the figure shows the fidelity between the states obtained after the first and the second projection onto $S$, the probability that the second projection is successful given that the first is successful, and the purity of the state after the second projection for two different values of $|\alpha|^2$.}
\end{figure}

We next investigate the possibility to use a comparison of the number of photons in the two modes to detect a loss. As in Sec.~\ref{Filter}, we start with the input state $|\psi_{\textrm{in}}\rangle=|\alpha\rangle|\alpha\rangle$ and use the proposed setup to project it onto the subspace $S$. We then use two beam splitters with reflectivity $R$ to model a fractional loss in both modes. After tracing out the reflected field, we finally use the proposed setup once more to project the state onto $S$. In Fig.~\ref{recover}, we plot the fidelity between the state obtained after the first projection onto $S$ and the state obtained after the second projection onto $S$, the probability that the second projection is successful given that the first is successful, and the purity
of the state after the second projection. The second projection is seen to recover the state obtained after the first projection with a fidelity close to unity even for losses of a few percent. This is the case because a loss of only one photon will always lead to a failure of the second conditional projection. The main contribution to the fidelity decrease for small $R$ is thus a simultaneous loss of one photon from both modes. It is also interesting to note that the final state is actually pure for all values of $R$, which is a consequence of the particular choice of input state. Finally, we note that a single unit is sufficient to detect loss of a single photon, and for small $R$ we thus only need to use one unit for the second projection in practice.

\begin{figure}
\includegraphics[width=\columnwidth]{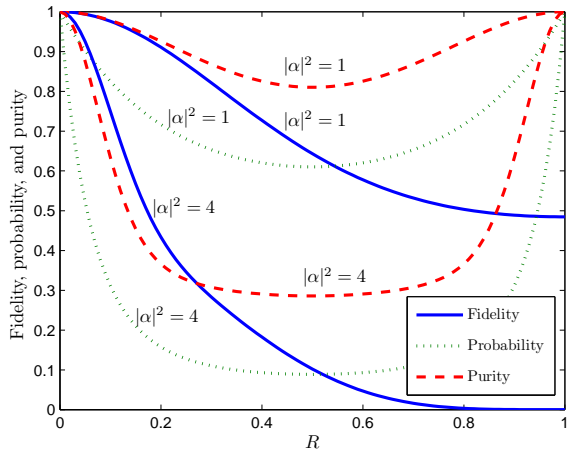}
\caption{\label{loss}(Color online) Projecting the input state $|\psi_{\textrm{in}}\rangle=|\alpha\rangle|\alpha\rangle|\alpha\rangle|\alpha\rangle$ onto the subspace spanned by the vectors $|n\rangle|n\rangle|n\rangle|n\rangle$, $n=0,1,2,\ldots$, followed by a fractional loss $R$ in all four modes and a projection of modes 1 and 2 onto $S$ and of modes 3 and 4 onto $S$, the figure shows the fidelity between the states obtained after the first and the second set of projections, the probability that the projections onto $S$ are successful given that the first projection is successful, and the purity of the state after the projections onto $S$ for two different values of $|\alpha|^2$.}
\end{figure}

Let us also consider a four mode input state $|\psi_{\textrm{in}}\rangle=|\alpha\rangle|\alpha\rangle|\alpha\rangle|\alpha\rangle$. As before we use the setup to project this state onto the subspace spanned by the state vectors $|n\rangle|n\rangle|n\rangle|n\rangle$, $n=0,1,2,\ldots$. We then imagine a fractional loss of $R$ to take place in all modes. If two of the modes are on their way to Alice and the two other modes are on their way to Bob, we can only try to recover the original projection by projecting the former two modes onto $S$ and the latter two modes onto $S$. The results are shown in Fig.~\ref{loss}, and again the curves showing the fidelity and the purity are seen to be very flat and close to unity for small losses. This scheme allows one to distribute entanglement with high fidelity, but reduced success probability.

\section{Deviations from ideal behavior}\label{Nonideal}

So far we have considered the ideal limit of infinitely long pulses and an infinitely strong coupling. In this section, we use a more detailed model of the interaction of the light field with an atom in a cavity to investigate how long the input pulses need to be and how large the single atom cooperativity parameter should be to approximately achieve this limit.

\subsection{Optimal input mode function}

The backreflection of light in a cavity leads to a state dependent distortion of the shape of the mode function of the input field, and to study this effect in more detail we concentrate on a single cavity as shown in Fig.~\ref{cavity} in the following. For simplicity, we assume the input field to be a continuous coherent state \cite{blow}
\begin{equation}\label{cont}
|\{\alpha_{\textrm{in}}(t)\}\rangle=\exp\left[\int\alpha_{\textrm{in}}(t)\hat{a}^\dag(t)dt
-\int\alpha_{\textrm{in}}^*(t)\hat{a}(t)dt\right]|0\rangle
\end{equation}
with mode function $f_{\textrm{in}}(t)=\alpha_{\textrm{in}}(t)/\alpha$, where $|\alpha|^2=\int|\alpha_{\textrm{in}}(t)|^2dt$ is the expectation value of the total number of photons in the input beam. The light-atom interaction is governed by the Hamiltonian
\begin{equation}
H=\hbar g(\hat{c}^\dag\sigma+\sigma^\dag\hat{c}),
\quad\sigma:=|{\uparrow}\rangle\langle e|
\end{equation}
and the decay term
\begin{equation}
\mathcal{L}\rho=\frac{\Gamma}{2}(2\sigma\rho\sigma^\dag-\sigma^\dag\sigma\rho-\rho\sigma^\dag\sigma),
\end{equation}
where $g$ is the light-atom coupling strength, $\hat{c}$ is the annihilation operator of the cavity field, $\Gamma$ is the decay rate of the excited state of the atom due to spontaneous emission, and $\rho$ is the density operator representing the state of the atom and the cavity field. For $g^2\textrm{Tr}(\hat{c}^\dag\hat{c}\langle{\uparrow}|\rho|{\uparrow}\rangle)\ll(\Gamma/2)^2$, where $\textrm{Tr}$ denotes the trace, the population in the excited state of the atom is very small, and we may eliminate this state adiabatically. This reduces the effective light-atom interaction dynamics to a single decay term
\begin{equation}
\frac{2g^2}{\Gamma}\left(2\hat{c}|{\uparrow}\rangle
\langle{\uparrow}|\rho|{\uparrow}\rangle\langle{\uparrow}|\hat{c}^\dag
-\hat{c}^\dag\hat{c}|{\uparrow}\rangle\langle{\uparrow}|\rho
-\rho|{\uparrow}\rangle\langle{\uparrow}|\hat{c}^\dag\hat{c}\right),
\end{equation}
i.e., the atom is equivalent to a beam splitter which reflects photons out of the cavity at the rate $4g^2/\Gamma$ if the atom is in the state $|{\uparrow}\rangle$ and does not affect the light field if the atom is in the state $|{\downarrow}\rangle$.

Assume the atom to be in the state $|j\rangle$, $j\in\{{\downarrow},{\uparrow}\}$. Since the input mirror of the cavity may be regarded as a beam splitter with high reflectivity, all components of the setup transform field operators linearly. For a coherent state input field, the cavity field and the output field are hence also coherent states. We may divide the time axis into small segments of width $\tau$ and approximate the integrals in \eqref{cont} by sums. The input state is then a direct product of single mode coherent states with amplitudes $\alpha_{\textrm{in}}(t_k)\sqrt{\tau}$ and annihilation operators $\hat{a}(t_k)\sqrt{\tau}$, where $t_k=k\tau$, $k=0,\pm1,\pm2,\ldots$. In the following, we choose $\tau$ to be equal to the round trip time of light in the cavity, which requires $\alpha_{\textrm{in}}(t)$ to vary slowly on that time scale. Denoting the coherent state amplitude of the cavity field at time $t$ by $\gamma_j(t)$, we use the beam splitter transformation
\begin{eqnarray}\label{bst}
\bigg[\textrm{out}\bigg]=\bigg[\begin{array}{cc}t_c&-r_c\\r_c&t_c\end{array}\bigg]
\bigg[\textrm{in}\bigg]
\end{eqnarray}
for the input mirror of the cavity to derive
\begin{eqnarray}
\gamma_j(t)&=&t_c\sqrt{\tau}\alpha_{\textrm{in}}(t)+r_ct_j\gamma_j(t-\tau),\label{gamma}\\
\sqrt{\tau}\alpha^{(j)}_{\textrm{out}}(t)&=&r_c\sqrt{\tau}
\alpha_{\textrm{in}}(t)-t_ct_j\gamma_j(t-\tau)\label{out},
\end{eqnarray}
where $r_c^2=1-t_c^2$ is the reflectivity of the input mirror of the cavity, $t_j^2$ is the transmissivity of the beam splitter, which models the loss due spontaneous emission, i.e., $t_j^2=1-4g^2\tau\delta_{j{\uparrow}}/\Gamma$, and $\alpha^{(j)}_{\textrm{out}}(t)$ denotes the output field from the cavity. We have here included an additional phase shift of $\pi$ per round trip in the cavity to ensure the input field to be on resonance with the cavity, i.e., the second element of the input vector in \eqref{bst} is $-t_j\gamma(t-\tau)$. Taking the limit $\tau\rightarrow0$ and $t_c^2\rightarrow0$ for fixed cavity decay rate $\kappa:=t_c^2/\tau$, \eqref{gamma} and \eqref{out} reduce to
\begin{eqnarray}
\frac{d\gamma_j(t)}{dt}&=&-\left(1+2C\delta_{j{\uparrow}}\right)\frac{\kappa}{2}\gamma_j(t)
+\sqrt{\kappa}\alpha_{\textrm{in}}(t),\label{gammaeq}\\
\alpha^{(j)}_{\textrm{out}}(t)&=&\alpha_{\textrm{in}}(t)-\sqrt{\kappa}\gamma_j(t)\label{inout},
\end{eqnarray}
where $C:=2g^2/(\kappa\Gamma)$ is the single atom cooperativity parameter. According to the steady state solution of \eqref{gammaeq}
\begin{equation}\label{ss}
\gamma_j(t)=\frac{1}{1+2C\delta_{j{\uparrow}}}\frac{2\alpha_{\textrm{in}}(t)}{\sqrt{\kappa}},
\end{equation}
we need $C\gg1$ to efficiently expel the light field from the cavity for $j={\uparrow}$. We should also remember the criterion for the validity of the adiabatic elimination, which by use of \eqref{ss} takes the form
\begin{equation}
1\gg\frac{4C}{(1+2C)^2}\frac{2|\alpha_{\textrm{in}}(t)|^2}{\Gamma}
\approx\frac{\kappa|\alpha_{\textrm{in}}(t)|^2}{g^2},
\end{equation}
i.e., for $C\gg1$ the average flux of photons in the input beam should not significantly exceed the average number of photons emitted spontaneously per unit time from an atom, which has an average probability of one half to be in the excited state.

Solving \eqref{gammaeq},
\begin{equation}\label{gammat}
\gamma_j(t)=\sqrt{\kappa}\int_{-\infty}^te^{-(1+2C\delta_{j{\uparrow}})
\kappa(t-t')/2}\alpha_{\textrm{in}}(t')dt',
\end{equation}
we obtain the output field $\alpha^{(j)}_{\textrm{out}}(t)=\alpha f^{(j)}_{\textrm{out}}(t)$ with
\begin{equation}\label{fout}
f^{(j)}_{\textrm{out}}(t)=f_{\textrm{in}}(t)-
\kappa\int_{-\infty}^te^{-(1+2C\delta_{j{\uparrow}})\kappa(t-t')/2}f_{\textrm{in}}(t')dt'.
\end{equation}
Note that $f^{(j)}_{\textrm{out}}(t)$ is not necessarily normalized due to the possibility of spontaneous emission. The ideal situation is $f^{(\uparrow)}_{\textrm{out}}(t)=f_{\textrm{in}}(t)$ and $f^{(\downarrow)}_{\textrm{out}}(t)=-f_{\textrm{in}}(t)$, and we would thus like the norm of
\begin{equation}
E_j:=1+(-1)^{\delta_{j{\uparrow}}}\int_{-\infty}^{\infty}f^*_{\textrm{in}}(t)
f^{(j)}_{\textrm{out}}(t)dt
\end{equation}
to be as small as possible.

For $j={\uparrow}$ and $C\gg1$, the exponential function in \eqref{fout} is practically zero unless $\kappa(t-t')\ll1$, and as long as $f_{\textrm{in}}(t)$ changes slowly on the time scale $(C\kappa)^{-1}$, we may take $f_{\textrm{in}}(t)$ outside the integral to obtain $f^{(\uparrow)}_{\textrm{out}}(t)=(1-2/(1+2C))f_{\textrm{in}}(t)$ and $E_{\uparrow}=2/(1+2C)\approx C^{-1}$. As this result is independent of $f_{\textrm{in}}(t)$, a natural criterion for the optimal choice of input mode function is to minimize
\begin{equation}
|E_{\downarrow}|=\left|2-\kappa\int_{-\infty}^{\infty}\int_{-\infty}^t
e^{-\kappa(t-t')/2}f^*_{\textrm{in}}(t)f_{\textrm{in}}(t')dt'dt\right|
\end{equation}
under the constraint $\int_{-\infty}^{\infty}f^*_{\textrm{in}}(t)f_{\textrm{in}}(t)dt=1$. We also restrict $f_{\textrm{in}}(t)$ to be zero everywhere outside the time interval $[-T/2,T/2]$. Since we would like the double integral to be as close to 2 as possible, we should choose $f^*_{\textrm{in}}(t)=f_{\textrm{in}}(t)$. A variational calculation then provides the optimal solution
\begin{equation}\label{mode}
f_{\textrm{in}}(t)=\left\{\begin{array}{cl}
A\cos(\omega_0t)&\textrm{for }t\in[-T/2,T/2]\\
0&\textrm{otherwise}
\end{array}
\right.,
\end{equation}
where
\begin{equation}
A=\left(\frac{2}{T+\sin(\omega_0T)/\omega_0}\right)^{1/2},
\end{equation}
\begin{equation}
\frac{2\omega_0}{\kappa}\tan\left(\frac{\omega_0T}{2}\right)=1,
\end{equation}
and $\omega_0T\in[0,\pi[$.

\begin{figure}
\includegraphics[width=\columnwidth]{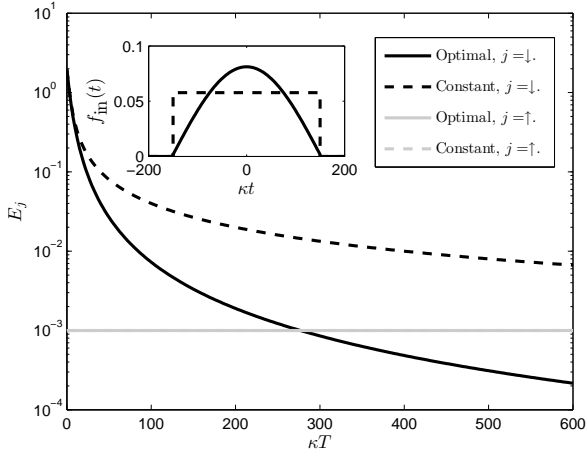}
\caption{\label{optmode}Deviation $E_j=1+(-1)^{\delta_{j\uparrow}}\int f^*_{\textrm{in}}(t)f^{(j)}_{\textrm{out}}(t)dt$ of the overlap between the output mode function and the input mode function from the ideal value when the atom is in the state $|j\rangle$. Note that $f^{(j)}_{\textrm{out}}(t)$ is defined such that the norm is less than unity if a loss occurs during the interaction. The single atom cooperativity parameter is assumed to be $C=10^3$. Solid lines are for the optimal input mode function given in \eqref{mode}, and dashed lines are for an input mode function, which is constant in the interval $t\in[-T/2,T/2]$ and zero otherwise. The inset illustrates these functions for $\kappa T=300$. For $j=\uparrow$, the shape of the output mode function is almost the same as the shape of the input mode function as long as $\kappa T\gg C^{-1}$, but the norm is slightly decreased due to spontaneous emission from the atom such that $E_{\uparrow}\approx C^{-1}$. This result is independent of the actual shape of the input mode function, and the solid and dashed lines for $E_{\uparrow}$ are hence indistinguishable in the figure. The results for $E_{\downarrow}$ are independent of $C$, because the light field does not interact with the atom in this case. Nonzero values of $E_{\downarrow}$ only occur due to distortion of the shape of the mode function. For $\kappa T\gg1$, $E_{\downarrow}\approx8\pi^2/(\kappa T)^2$ for the optimal input mode function and $E_{\downarrow}\approx4/(\kappa T)$ for the constant mode function.}
\end{figure}

This solution gives
\begin{equation}
E_{\downarrow}=\frac{2x^2}{1+x^2}=2\cos^2\left(\frac{\omega_0T}{2}\right),\quad x:=\frac{2\omega_0}{\kappa}.
\end{equation}
For fixed $\kappa$, $\omega_0$ is a decreasing function of $T$, and for $\kappa T\gg2\pi$, $\omega_0\approx\pi/T$ and $E_{\downarrow}\approx8\pi^2/(\kappa T)^2$. Distortion of the input mode function can thus be avoided by choosing a sufficiently long input pulse. This may be understood by considering the Fourier transform of the input mode function. For a very short pulse the frequency distribution is very broad and only a small part of the field is actually on resonance with the cavity, i.e., most of the field is reflected without entering into the cavity. For a very long pulse, on the other hand, the frequency distribution is very narrow and centered at the resonance frequency of the cavity. $E_j$ is plotted in Fig.~\ref{optmode} as a function of $T$ both for the optimal input mode function and an input mode function, which is constant in the interval $[-T/2,T/2]$. The optimal input mode function is seen to provide a significant improvement. In fact, for the constant input mode function $E_{\downarrow}=4[1-\exp(-\kappa T/2)]/(\kappa T)$, and hence $E_{\downarrow}$ scales only as $(\kappa T)^{-1}$ for large $\kappa T$. Finally, we note that $\tau=t_c^2/\kappa<\kappa^{-1}$, and $\kappa T\gg1$ thus also ensures $\tau/T\ll1$, which justifies the approximation in Eqs.~\ref{gammaeq} and \ref{inout}.

\subsection{Single atom cooperativity parameter}\label{cooperativity}

We would like to determine how a single unit of the setup transforms the state of the field, when we use the full multi-mode description. For this purpose it is simpler to work in frequency space, and we thus use the definition of the Fourier transform $f(\omega)=\int f(t)\exp(-i\omega t)dt/\sqrt{2\pi}$ on Eqs.~\eqref{gammat} and \eqref{fout} to obtain
\begin{eqnarray}
\gamma_j(\omega)&=&\frac{\sqrt{\kappa}}{(1+2C\delta_{j{\uparrow}})\kappa/2+i\omega}
\alpha_{\textrm{in}}(\omega),\label{gammao}\\
\alpha^{(j)}_{\textrm{out}}(\omega)&=&K_j(\omega)\alpha_{\textrm{in}}(\omega),\label{tfield}
\end{eqnarray}
where
\begin{equation}\label{Kj}
K_j(\omega):=-\frac{(1-2C\delta_{j{\uparrow}})\kappa/2-i\omega}
{(1+2C\delta_{j{\uparrow}})\kappa/2+i\omega}.
\end{equation}
Assume now that the density operator of the two-beam input field to the unit may be written on the form
\begin{multline}\label{rhoin}
\rho_{\textrm{in}}=\sum_n\sum_mc_{nm}|\{\alpha_n(\omega)\}\rangle\langle\{\alpha_m(\omega)\}|\\
\otimes|\{\beta_n(\omega)\}\rangle\langle\{\beta_m(\omega)\}|,
\end{multline}
where $n$ and $m$ are summed over the same set of numbers and $|\{\alpha_n(\omega)\}\rangle$ and $|\{\beta_n(\omega)\}\rangle$ are continuous coherent states in frequency space, i.e.,
\begin{multline}\label{ccs}
|\{\alpha_n(\omega)\}\rangle=\exp\bigg[\int\alpha_n(\omega)\hat{a}^\dag(\omega)d\omega\\
-\int\alpha_n^*(\omega)\hat{a}(\omega)d\omega\bigg]|0\rangle
\end{multline}
and similarly for $|\{\beta_n(\omega)\}\rangle$. Note that \eqref{ccs} and \eqref{cont} are consistent when $\alpha_n(\omega)$ and $\alpha_n(t)$ as well as $\hat{a}(\omega)$ and $\hat{a}(t)$ are related through a Fourier transform \cite{blow}. After the first 50:50 beam splitter, the input state is transformed into
\begin{multline}
\rho'=\sum_n\sum_mc_{nm}
\left|\left\{(\alpha_n(\omega)+\beta_n(\omega))/\sqrt{2}\right\}\right\rangle\\
\left\langle\left\{(\alpha_m(\omega)+\beta_m(\omega))/\sqrt{2}\right\}\right|\\
\otimes\left|\left\{(\beta_n(\omega)-\alpha_n(\omega))/\sqrt{2}\right\}\right\rangle\\
\left\langle\left\{(\beta_m(\omega)-\alpha_m(\omega))/\sqrt{2}\right\}\right|,
\end{multline}
and this is the input state to the two cavities. The initial state of the two atoms is
\begin{equation}
\rho_{\textrm{at}}=\frac{1}{4}\sum_{i\in\{\downarrow,\uparrow\}}
\sum_{j\in\{\downarrow,\uparrow\}}\sum_{p\in\{\downarrow,\uparrow\}}
\sum_{q\in\{\downarrow,\uparrow\}}
|i\rangle\langle p|\otimes|j\rangle\langle q|.
\end{equation}
We thus need to know how one of the cavities transforms a term like $|\alpha_{\textrm{in},n}(\omega)\rangle\langle\alpha_{\textrm{in},m}(\omega)|\otimes|i\rangle\langle p|$.

We saw in the last subsection that a cavity is equivalent to an infinite number of beam splitter operations applied to the cavity mode and the input field modes. To take the possibility of spontaneous emission into account, we also apply a beam splitter operation to the cavity mode and a vacuum mode in each time step and subsequently trace out the vacuum mode. As the beam splitters are unitary operators acting from the left and the right on the density operator, the field amplitudes are transformed according to \eqref{tfield} as before, but when the ket and the bra are different, the trace operations lead to a scalar factor. Since the reflectivity of the beam splitter modeling the loss is $2C\kappa\tau\delta_{j\uparrow}$ for an atom in the state $|j\rangle$, this factor is
\begin{multline}
d_{ip}[\alpha_{\textrm{in},n}(\omega),\alpha_{\textrm{in},m}(\omega)]\\
=\prod_{k=-\infty}^\infty\langle\sqrt{2C\kappa\tau}\gamma_{m,p}(k\tau)\delta_{p\uparrow}|
\sqrt{2C\kappa\tau}\gamma_{n,i}(\kappa\tau)\delta_{i\uparrow}\rangle\\
=\exp\bigg\{-\int_{-\infty}^{\infty}\frac{C\kappa^2}{(1+2C)^2(\kappa/2)^2+\omega^2}
\big[|\alpha_{\textrm{in},n}(\omega)|^2\delta_{i\uparrow}\\
+|\alpha_{\textrm{in},m}(\omega)|^2\delta_{p\uparrow}
-2\alpha_{\textrm{in},n}(\omega)\alpha^*_{\textrm{in},m}
(\omega)\delta_{i\uparrow}\delta_{p\uparrow}\big]d\omega\bigg\},
\end{multline}
where $\gamma_{n,i}(t)$ is the amplitude of the cavity field corresponding to the input field $\alpha_{\textrm{in},n}(t)$ and the atomic state $|i\rangle$ as given in Eq.~\eqref{gammat}. Altogether, $|\alpha_{\textrm{in},n}(\omega)\rangle\langle\alpha_{\textrm{in},m}(\omega)|
\otimes|i\rangle\langle p|$ is thus transformed into
$d_{ip}[\alpha_{\textrm{in},n}(\omega),\alpha_{\textrm{in},m}(\omega)]
|K_i(\omega)\alpha_{\textrm{in},n}(\omega)\rangle
\langle K_p(\omega)\alpha_{\textrm{in},m}(\omega)|
\otimes|i\rangle\langle p|$. Projecting both atoms onto $(|{\uparrow}\rangle+|{\downarrow}\rangle)/2$ and taking the final 50:50 beam splitter into account, we obtain the output state after one unit
\begin{widetext}
\begin{multline}\label{rhoout}
\rho_{\textrm{out}}=\sum_n\sum_m\sum_{i\in\{\downarrow,\uparrow\}}\sum_{j\in\{\downarrow,\uparrow\}}
\sum_{p\in\{\downarrow,\uparrow\}}\sum_{q\in\{\downarrow,\uparrow\}}\frac{1}{16}c_{nm}
d_{ip}[(\alpha_n(\omega)+\beta_n(\omega))/\sqrt{2},(\alpha_m(\omega)+\beta_m(\omega))/\sqrt{2}]\\
\times d_{jq}[(\beta_n(\omega)-\alpha_n(\omega))/\sqrt{2},(\beta_m(\omega)-\alpha_m(\omega))/\sqrt{2}]\\
\times\left|\left\{\frac{1}{2}(K_i+K_j)\alpha_n(\omega)
+\frac{1}{2}(K_i-K_j)\beta_n(\omega)\right\}\right\rangle
\left\langle\left\{\frac{1}{2}(K_p+K_q)\alpha_m(\omega)
+\frac{1}{2}(K_p-K_q)\beta_m(\omega)\right\}\right|\\
\otimes\left|\left\{\frac{1}{2}(K_i+K_j)\beta_n(\omega)
+\frac{1}{2}(K_i-K_j)\alpha_n(\omega)\right\}\right\rangle
\left\langle\left\{\frac{1}{2}(K_p+K_q)\beta_m(\omega)
+\frac{1}{2}(K_p-K_q)\alpha_m(\omega)\right\}\right|,
\end{multline}
\end{widetext}
where we have written $K_i=K_i(\omega)$ for brevity. A subsequent phase shifter is easily taken into account by multiplying the coherent state amplitudes by the appropriate phase factors. We note that the result has the same form as the input state \eqref{rhoin} if we collect $n$, $i$, and $j$ into one index and $m$, $p$, and $q$ into another index. We can thus apply the transformation repeatedly to obtain the output state after $N+1$ units of the setup.

\begin{figure}[b]
\hspace{0.15\textwidth}
\includegraphics[width=\columnwidth]{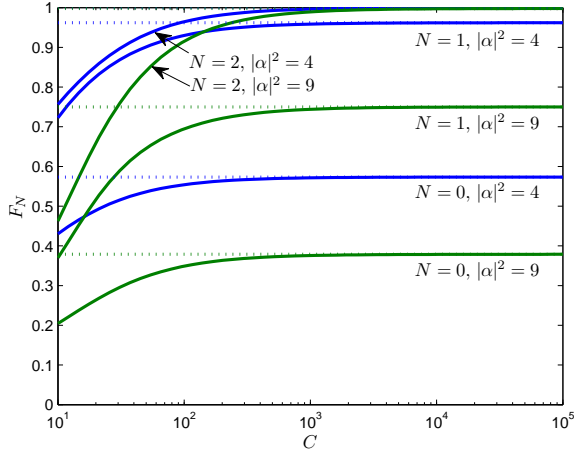}
\caption{\label{finiteC}(Color online) Fidelity as defined in \eqref{FN} as a function of the single atom cooperativity parameter $C$ for different values of the expectation value of the number of photons in one of the input modes and different numbers of units of the setup, assuming $\kappa T\gg1$ and $\rho_{\textrm{in}}=|\alpha\rangle\langle\alpha|\otimes|\alpha\rangle\langle\alpha|$. The dotted lines are the asymptotes for $C\rightarrow\infty$.}
\end{figure}

If we assume $\alpha_n(\omega)$ and $\beta_m(\omega)$ to have the same shape for all $n$ and $m$, the output field simplifies to a two-mode state for $\kappa T\gg1$ as expected. This may be seen as follows. When $T$ is much larger than $\kappa^{-1}$, the width of the distribution $\alpha_n(\omega)$ in frequency space is much smaller than $\kappa$. For the relevant frequencies we thus have $\omega\ll\kappa/2$, and in this case the right hand side of \eqref{Kj} reduces to $-(1-2C\delta_{j{\uparrow}})/(1+2C\delta_{j{\uparrow}})$, i.e., $K_j$ is independent of frequency, and the amplitudes of all the continuous coherent states of the output density operator are proportional to $\alpha_n(\omega)$.

In Fig.~\ref{finiteC}, we have used \eqref{rhoout} in the limit $\kappa T\gg1$ to compute the fidelity \eqref{FN} as a function of the single atom cooperativity parameter for $\rho_{\textrm{in}}=|\alpha\rangle\langle\alpha|\otimes|\alpha\rangle\langle\alpha|$. The ideal values for $C\rightarrow\infty$ are also shown, and the deviations are seen to be very small for $C$ larger than about $10^3$. Current experiments have demonstrated single atom cooperativity parameters on the order of $10^2$ \cite{strongcoupling1,strongcoupling2,strongcoupling3}, and high fidelities can also be achieved for this value. When $C$ is finite, there is a relative photon loss of $1-\int|f^{({\uparrow})}_\textrm{out}(t)|^2dt\approx2/C$ due to spontaneous emission each time the field interacts with a cavity containing an atom in the state $|{\uparrow}\rangle$. We note, however, that the setup is partially robust against such losses because the next unit of the setup removes all components of the state for which a single photon has been lost.

\section{Conclusion}\label{Conclusion}

The setup put forward and studied in this article acts in many respects like a photon number filter and has several attractive applications for quantum technologies. Based on the ability to distinguish even and odd photon numbers using the interaction of the light field with a high finesse optical cavity, photonic two mode input states can be projected onto photon-number correlated states. Naturally, this protocol is very well suited to detect losses and can in particular be adapted to purify photon-number entangled states in quantum communication. We studied deviations from ideal behavior such as finite length of the input pulses and limited coupling to estimate for which parameters the idealized description is valid. The setup can be modified such that it is capable to perform a quantum-non-demolition measurement of photon numbers in the optical regime. The non-destructive photon counting device completes the versatile toolbox provided by the proposed scheme.

\begin{acknowledgments}
We acknowledge valuable discussions with Ignacio Cirac and Stefan Kuhr and support from the Danish Ministry of Science, Technology, and Innovation, the Elite Network of Bavaria (ENB) project QCCC, the EU projects SCALA, QAP, COMPAS, the DFG-Forschungsgruppe 635, and the DFG Excellence Cluster MAP.
\end{acknowledgments}


\begin{thebibliography}{99}

\bibitem{strongcoupling1} H. J. Kimble, Phys. Scr. {\bf T76}, 127 (1998).

\bibitem{strongcoupling2} A. Boca, R. Miller, K. M. Birnbaum, A. D. Boozer, J. McKeever, and H. J. Kimble, Phys. Rev. Lett. {\bf93}, 233603 (2004).

\bibitem{strongcoupling3} Y. Colombe, T. Steinmetz, G. Dubois, F. Linke, D. Hunger, and J. Reichel, Nature (London) {\bf450}, 272 (2007).

\bibitem{strongcoupling4} K. M. Birnbaum, A. Boca, R. Miller, A. D. Boozer, T. E. Northup, and H. J. Kimble, Nature (London) \textbf{436}, 87 (2005).

\bibitem{HongOuMandel} C. K. Hong, Z. Y. Ou, and L. Mandel, Phys. Rev. Lett. {\bf59}, 2044 (1987).

\bibitem{EntanglementFilter} R. Okamoto, J. L. O'Brien, H. F. Hofmann, T. Nagata, K. Sasaki, and S. Takeuchi, Science {\bf323}, 483 (2009).

\bibitem{EPR} A. Einstein, B. Podolsky, and N. Rosen, Phys. Rev. {\bf47}, 777 (1935).

\bibitem{Teleportation1} C. H. Bennett, G. Brassard, C. Cr\'epeau, R. Jozsa, A. Peres, and W. K. Wootters, Phys. Rev. Lett. {\bf70}, 1895 (1993).

\bibitem{Teleportation2} A. Furusawa, J. L. S{\o}rensen, S. L. Braunstein, C. A. Fuchs, H. J. Kimble, and E. S. Polzik, Science {\bf282}, 706 (1998).

\bibitem{Teleportation3} N. Takei, H. Yonezawa, T. Aoki, and A. Furusawa, Phys. Rev. Lett. {\bf94}, 220502 (2005).

\bibitem{Teleportation4} L. Vaidman, Phys. Rev. A {\bf49}, 1473 (1994).

\bibitem{Teleportation5} S. L. Braunstein and H. J. Kimble, Phys. Rev. Lett. {\bf80}, 869 (1998).

\bibitem{Swapping1} M. \.Zukowski, A. Zeilinger, M. A. Horne, and A. K. Ekert, Phys. Rev. Lett. {\bf71}, 4287 (1993).

\bibitem{Swapping2} P. van Loock and S. L. Braunstein, Phys. Rev. A {\bf61}, 010302(R) (1999).

\bibitem{Swapping3} X. Jia, X. Su, Q. Pan, J. Gao, C. Xie, and K. Peng, Phys. Rev. Lett. {\bf93}, 250503 (2004).

\bibitem{QKD1} C. H. Bennett and G. Brassard, in Proceedings of IEEE International Conference on Computers, Systems and Signal Processing, Bangalore, India (IEEE, New York, 1984), pp 175-179.

\bibitem{QKD2} S. L. Braunstein and P. van Loock, Rev. Mod. Phys. {\bf77}, 513 (2005).

\bibitem{Bell1} J. S. Bell, Physics {\bf1}, 195 (1964).

\bibitem{Bell2} E. G. Cavalcanti, C. J. Foster, M. D. Reid, and P. D. Drummond, Phys. Rev. Lett. {\bf99}, 210405 (2007).


\bibitem{Purification1} J.-W. Pan, S. Gasparoni, R. Ursin, G. Weihs, and A. Zeilinger, Nature {\bf423}, 417 (2003).

\bibitem{Purification2} Z. Zhao, T. Yang, Y.-A. Chen, A.-N. Zhang, and J.-W. Pan, Phys. Rev. Lett. {\bf90}, 207901 (2003).


\bibitem{Purification3} B. Hage, A. Samblowski, J. DiGuglielmo, A. Franzen, J. Fiur\'a\v sek, and R. Schnabel, Nature Phys. {\bf4}, 915 (2008).

\bibitem{Purification4} R. Dong, M. Lassen, J. Heersink, C. Marquardt, R. Filip, G. Leuchs, U. L. Andersen, Nature Phys. {\bf4}, 919 (2008).


\bibitem{GaussianImp1} G. Giedke and J. I. Cirac, Phys. Rev. A {\bf66}, 032316 (2002).

\bibitem{GaussianImp2} J. Eisert, S. Scheel, and M. B. Plenio, Phys. Rev. Lett. {\bf89}, 137903 (2002).

\bibitem{GaussianImp3} J. Fiur\'a\v sek, Phys. Rev. Lett. {\bf89}, 137904 (2002).


\bibitem{PurificationProp1} J. Eisert, D. Browne, S. Scheel, and M. Plenio, Ann. Phys. (N.Y.) \textbf{311}, 431 (2004).

\bibitem{PurificationProp2} D. Menzies and N. Korolkova, Phys. Rev. A \textbf{76}, 062310 (2007).

\bibitem{PurificationProp3} A. Ourjoumtsev, A. Dantan, R. Tualle-Brouri, and P. Grangier, Phys. Rev. Lett. {\bf98}, 030502 (2007).

\bibitem{PurificationProp4} T. Opatrn\'y, G. Kurizki, and D.-G. Welsch, Phys. Rev. A, {\bf61}, 032302 (2000).

\bibitem{PurificationProp5} L.-M. Duan, G. Giedke, J. I. Cirac, and P. Zoller, Phys. Rev. Lett {\bf84}, 4002 (2000).

\bibitem{PurificationProp6} J. Fiur\'a\v sek, Jr. L. Mi\v sta, and R. Filip, Phys. Rev. A {\bf67}, 022304 (2003).

\bibitem{jonas} H. Takahashi, J. S. Neergaard-Nielsen, M. Takeuchi, M. Takeoka, K. Hayasaka, A. Furusawa, and M. Sasaki, Nature Photonics {\bf4}, 178 (2010).


\bibitem{KLM} E. Knill, R. Laflamme, and G. J. Milburn, Nature {\bf409}, 46 (2001).

\bibitem{Kok07} P. Kok, W. J. Munro, K. Nemoto, T. C. Ralph, J. P. Dowling, and G. J. Milburn, Rev. Mod. Phys. {\bf79}, 135 (2007).

\bibitem{Obrian07} J. L. O'Brien, Science {\bf318}, 1567 (2007).


\bibitem{Prep1} C. \'Sliwa and K. Banaszek, Phys. Rev. A {\bf67}, 030101(R) (2003).

\bibitem{Prep2} T. B. Pittman, M. M. Donegan, M. J. Fitch, B. C. Jacobs, J. D. Franson, P. Kok, H. Lee, and J. P. Dowling, IEEE J. Sel. Top. Quant. Elec. {\bf9}, 1478 (2003).

\bibitem{Prep3} P. van Loock and N. L\"utkenhaus, Phys. Rev. A {\bf69}, 012302 (2004).


\bibitem{Zukowski93} M. \.Zukowski, A. Zeilinger, M. A. Horne, and A. K. Ekert, Phys. Rev. Lett. {\bf71}, 4287 (1993).


\bibitem{Crypto1} W.-Y. Hwang, Phys. Rev. Lett. {\bf91}, 057901 (2003).

\bibitem{Crypto2} J. Calsamiglia, S. M. Barnett, and N. L\"utkenhaus, Phys. Rev. A. {\bf65}, 012312 (2001).


\bibitem{Interferometry} K. Edamatsu, R. Shimizu, and T. Itoh, Phys. Rev. Lett. {\bf89}, 213601 (2002).



\bibitem{LightSources1} A. J. Shields, Nature Photonics {\bf1}, 215 (2007).

\bibitem{LightSources2} D. Achilles, C. Silberhorn, and I. A. Walmsley, Phys. Rev. Lett. {\bf97}, 043602 (2006).



\bibitem{APD} B. E. Kardyna\l, Z. L. Yuan, and A. J. Shields, Nature Photonics {\bf2}, 425 (2008).

\bibitem{Silberhorn04} D. Achilles, C. Silberhorn, C. Sliwa, K. Banaszek, I. A. Walmsley, M. J. Fitch, B. C. Jacobs, T. B. Pittman, and J. D. Franson, J. Mod. Opt. {\bf51}, 1499 (2004).

\bibitem{Banzek03} K. Banaszek and I. A. Walmsley, Optics Lett. {\bf28}, 52 (2003).

\bibitem{Cryo1} J. Kim, S. Takeuchi, S. Yamamoto, and H. H. Hogue, Appl. Phys. Lett. {\bf74}, 902 (1999).

\bibitem{Cryo2} B. Cabrera, R. M. Clarke, P. Colling, A. J. Miller, S. Nam, and R. W. Romani, Appl. Phys. Lett. {\bf73}, 735 (1998).

\bibitem{Cryo3} S. Somani, S. Kasapi, K. Wilsher, W. Lo, R. Sobolewski, and G. N. Gol'stman, Vac. Sci. Technol. B {\bf19}, 2766 (2001).

\bibitem{Others1} P. Eraerds, M. Legr\'e, J. Zhang, H. Zbinden, and N. Gisin, Journal of Lightwave Technology {\bf28}, 952 (2010).

\bibitem{Others2} A. J. Miller, S. W. Nam, J. M. Martinis, and A. V. Sergienko, Appl. Phys. Lett. {\bf83}, 791 (2003).

\bibitem{Others3} M. Fujiwara and M. Sasaki, Appl. Phys. Lett. {\bf86}, 111119 (2005).

\bibitem{QuantumDots1} B. E. Kardyna\l, S. S. Hees, A. J. Shields, C. Nicoll, I. Farrer, and D. A. Ritchie, Appl. Phys. Lett. {\bf90}, 181114 (2007).

\bibitem{QuantumDots2} E. J. Gansen, M. A. Rowe, M. B. Greene, D. Rosenberg, T. E. Harvey, M. Y. Su, R. H. Hadfield, S. W. Nam, and R. P. Mirin, Nature Photon. {\bf1}, 585 (2007).

\bibitem{QuantumJumps} S. Gleyzes, S. Kuhr, C. Guerlin, J. Bernu, S. Del\'eglise, U. B. Hoff, M. Brune, J.-M. Raimond, and S. Haroche, Nature (London) {\bf446}, 297 (2007).

\bibitem{QND1} C. Guerlin, J. Bernu, S. Del\'eglise, C. Sayrin, S. Gleyzes, S. Kuhr, M. Brune, J.-M. Raimond, and S. Haroche, Nature (London) {\bf448}, 889 (2007).

\bibitem{QND2} D. I. Schuster, A. A. Houck, J. A. Schreier, A. Wallraff, J. M. Gambetta, A. Blais, L. Frunzio, J. Majer, B. Johnson, M. H. Devoret, S. M. Girvin, and R. J. Schoelkopf, Nature (London) {\bf445}, 515 (2007).


\bibitem{QNDprop1} N. Imoto, H. A. Haus, and Y. Yamamoto, Phys. Rev. A \textbf{32}, 2287 (1985).

\bibitem{QNDprop2} M. J. Holland, D. F. Walls, and P. Zoller, Phys. Rev. Lett. \textbf{67}, 1716 (1991).

\bibitem{QNDprop3}  K. Jacobs, P. Tombesi, M. J. Collett, and D. F. Walls, Phys. Rev. A \textbf{49}, 1961 (1994).

\bibitem{QNDprop4} W. J. Munro, K. Nemoto, R. G. Beausoleil, and T. P. Spiller,  Phys. Rev. A \textbf{71}, 033819 (2005).

\bibitem{QNDprop5} J. Larson and M. Abdel-Aty,  Phys. Rev. A \textbf{80}, 053609 (2009).

\bibitem{QNDprop6} T. Bastin, J von Zanthier, and E. Solano, J. Phys. B: At. Mol. Opt. Phys. \textbf{39}, 685 (2006).


\bibitem{interaction} L.-M. Duan and H. J. Kimble, Phys. Rev. Lett. {\bf92}, 127902 (2004).

\bibitem{cat} B. Wang and L.-M. Duan, Phys. Rev. A {\bf72}, 022320 (2005).

\bibitem{parity} J. Kerckhoff, L. Bouten, A. Silberfarb, and H. Mabuchi, Phys. Rev. A {\bf79}, 024305 (2009).

\bibitem{switch} H. Mabuchi, Phys. Rev. A {\bf80}, 045802 (2009).

\bibitem{reichelreadout} R. Gehr, J. Volz, G. Dubois, T. Steinmetz, Y. Colombe, B. L. Lev, R. Long, J. Est\`eve, and J. Reichel, arXiv:1002.4424.

\bibitem{rempereadout} J. Bochmann, M. M\"ucke, C. Guhl, S. Ritter, G. Rempe, and D. L. Moehring, arXiv:1002.2918.

\bibitem{circuitQED} R. J. Schoelkopf and S. M. Girvin, Nature (London) {\bf451}, 664 (2008).

\bibitem{NOON} A. N. Boto, P. Kok, D. S. Abrams, S. L. Braunstein, C. P. Williams, and J. P. Dowling, Phys. Rev. Lett. {\bf85}, 2733 (2000).

\bibitem{blow} K. J. Blow, R. Loudon, S. J. D. Phoenix, and T. J. Shepherd, Phys. Rev. A {\bf42}, 4102 (1990).

\end{thebibliography}
\end{document}